\documentclass[conference, onecolumn]{IEEEtran}
\usepackage{listings}
\usepackage{graphicx}
\usepackage{subcaption}
\usepackage[table,xcdraw]{xcolor}
\usepackage{amsmath}
\usepackage{pifont}
\usepackage{amssymb}
\definecolor{mydarkblue}{rgb}{0,0.08,0.65}
\usepackage[colorlinks=true,
    linkcolor=mydarkblue,
    citecolor=mydarkblue,
    filecolor=mydarkblue,
    urlcolor=mydarkblue]{hyperref} 
\usepackage{cleveref}
\usepackage{soul}
\usepackage[shortlabels]{enumitem}
\usepackage{url}
\usepackage{color}
\usepackage{tikz}
\usepackage{balance}
\usepackage{multirow}
\usepackage{comment}
\usepackage{wrapfig,lipsum,booktabs}
\usepackage[frozencache,cachedir=.]{minted}

\usepackage[symbol]{footmisc}
\usepackage{tablefootnote}
\usepackage{tabularx}
\usepackage{placeins}
\usepackage{algorithm}
\usepackage{algpseudocode}
\usepackage{booktabs}
\usepackage{subcaption}

\usepackage{cuted}
\usepackage{float}
\usepackage{caption}

\usepackage{multicol}

\usepackage{dblfloatfix}
\usepackage{natbib}

\definecolor{codegreen}{rgb}{0,0.6,0}
\definecolor{codegray}{rgb}{0.5,0.5,0.5}
\definecolor{codepurple}{rgb}{0.58,0,0.82}
\definecolor{backcolour}{rgb}{0.95,0.95,0.92}

\makeatletter
\def\blfootnote{\xdef\@thefnmark{}\@footnotetext}
\makeatother

\lstdefinestyle{mystyle}{
  backgroundcolor=\color{backcolour},   commentstyle=\color{codegreen},
  keywordstyle=\color{magenta},
  numberstyle=\tiny\color{codegray},
  stringstyle=\color{codepurple},
  basicstyle=\ttfamily\footnotesize,
  breakatwhitespace=false,         
  breaklines=true,                 
  captionpos=b,                    
  keepspaces=true,                 
  numbers=left,                    
  numbersep=5pt,                  
  showspaces=false,                
  showstringspaces=false,
  showtabs=false,                  
  tabsize=2,
}

\AtBeginDocument{%
  \providecommand\BibTeX{{%
    \normalfont B\kern-0.5em{\scshape i\kern-0.25em b}\kern-0.8em\TeX}}}

\IEEEoverridecommandlockouts
\begin{document}

\title{ZUNA: Flexible EEG Superresolution with Position-Aware Diffusion Autoencoders}

\newcommand{\corr}{\textsuperscript{*}}

\author{
\IEEEauthorblockN{ 
Christopher Warner\textsuperscript{1,$\dagger$,*},
Jonas Mago\textsuperscript{1,$\dagger$},
JR Huml\textsuperscript{1},
Mohamed Osman\textsuperscript{1},
Beren Millidge\textsuperscript{1,*}
}
\IEEEauthorblockA{\textsuperscript{1}Zyphra}
\IEEEauthorblockA{\textsuperscript{$\dagger$}Joint First Authors}
\IEEEauthorblockA{\textsuperscript{*}Corresponding authors: \texttt{chris@zyphra.com}, \texttt{beren@zyphra.com}}
}
\maketitle

\setcounter{page}{1}

\begin{abstract}
We present \texttt{ZUNA}, a 380M-parameter masked diffusion autoencoder trained to perform masked channel infilling and superresolution for arbitrary electrode numbers and positions in EEG signals. The \texttt{ZUNA} architecture tokenizes multichannel EEG into short temporal windows and injects spatiotemporal structure via a 4D rotary positional encoding over (x,y,z,t), enabling inference on arbitrary channel subsets and positions. We train ZUNA on an aggregated and harmonized corpus spanning 208 public datasets containing approximately 2 million channel-hours using a combined reconstruction and heavy channel-dropout objective. We show that \texttt{ZUNA} substantially improves over ubiquitous spherical-spline interpolation methods, with the gap widening at higher dropout rates. Crucially, compared to other deep learning methods in this space, \texttt{ZUNA}'s performance \emph{generalizes} across datasets and channel positions allowing it to be applied directly to novel datasets and problems. Despite its generative capabilities, \texttt{ZUNA} remains computationally practical for deployment. We release Apache-2.0 weights and an MNE-compatible preprocessing/inference stack to encourage reproducible comparisons and downstream use in EEG analysis pipelines.
\end{abstract}

\section{Introduction}

\begin{figure}
    \centering
    \includegraphics[width=0.9\linewidth]{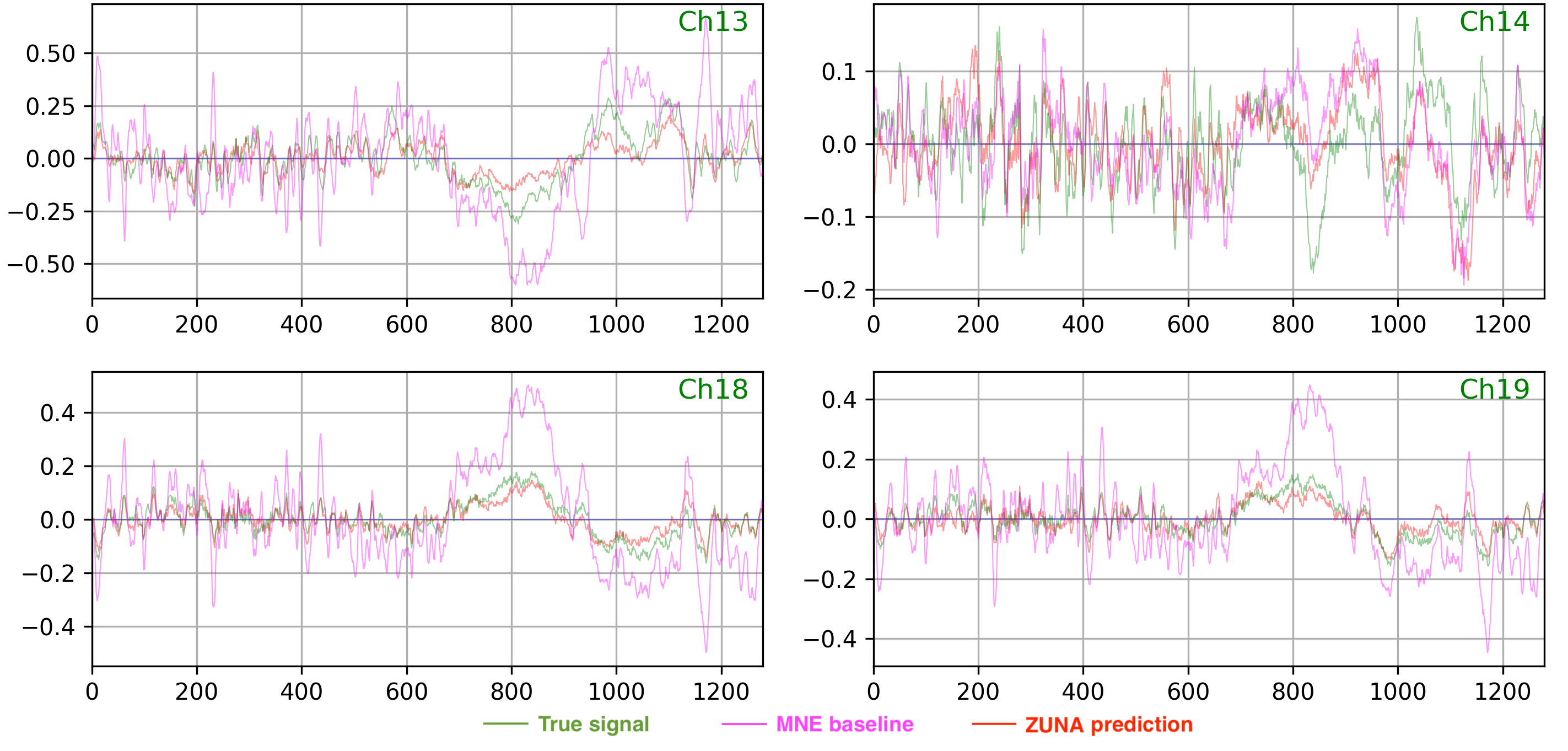}
    \caption{Examples of dropped-out EEG signals and their reconstruction by \texttt{ZUNA} and \texttt{MNE spherical-spline interpolation}. Green: ground-truth signal dropped-out (not presented to the model). Red: ZUNA model reconstruction. Magenta: MNE baseline spherical-spline interpolation.}
    \label{fig:Reconstructions}
\end{figure}

Electroencephalography (EEG) is the practice of recording electrical signals from either the scalp or the surface of the cortex. Scalp EEG is widely used in clinical and research settings due to its non-invasive nature and relatively low cost. Despite representing the averaged spiking rates of many billions of neurons, EEG signals still contain large amounts of useful and extractable information about brain function. EEG measurements have been widely applied in clinical settings to diagnose and study disorders such as epilepsy \citep{smith2005eeg, acharya2013automated, noachtar2009role, tan2025automatic} and multiple sclerosis \citep{mouazen2025machine, striano2003epileptic}.  EEG analysis can also be used to determine the stages of sleep \citep{supratak2017deepsleepnet, zhang2024review, li2022deep}, diagnose sleep disorders \citep{tiwari2022supervised}, and have been the basis of our understanding of various fundamental brain rhythms such as alpha, beta, gamma, and theta waves within the brain \citep{buzsaki2006rhythms}. More broadly, EEG signals can be used to detect and classify meditative states \citep{schoenberg2019mapping, kora2021eeg, faber2017eeg}, measure attentional focus \citep{su2022stanet, al2022predict, kaushik2022decoding}, and can even be used to extract some degree of visual, linguistic, or motor information from the cortex \citep{song2023decoding, wilson2024feasibility, tam2019human, alazrai2019deep, jerbi2011inferring}.

Since EEG data can be collected non-invasively using relatively simple, portable and affordable devices including, as of recently, consumer-grade headsets and headbands \citep{garcia2025eeg, biondi2022noninvasive}, EEG signals provide perhaps the most promising path to widely deployed brain-computer-interface (BCI) technology. If sufficient information exists in the EEG signal and can be decoded, then it  becomes possible to design headsets and other wearables that enable a higher bandwidth interaction with a computer system than voice or text or GUI interactions. \citep{varbu2022past, khademi2023review, elashmawi2024comprehensive}

\subsection{Motivation}

The challenge however is that the information about mental states that is encoded and accessible within scalp EEG signals is still uncertain, as is the temporal and spatial resolution needed to decode it. Researchers have successfully decoded global brainstates such as focus \citep{su2022stanet, al2022predict, kaushik2022decoding}, emotions \citep{liang2019unsupervised, duan2020machine}, meditative states \citep{schoenberg2019mapping, kora2021eeg, faber2017eeg}, stages of sleep \citep{supratak2017deepsleepnet, zhang2024review, li2022deep} and many more from EEG data using relatively simple methods. Moreover, researchers have harnessed pretrained neural networks to additionally decode specific stimuli such as images \citep{song2023decoding, wilson2024feasibility, xu2025alljoined}, text \citep{zhang2018converting, murad2024unveiling, wang2024enhancing}, and motor plans \citep{jerbi2011inferring, muller2016classic, omedes2018factors, wang2020enhance} from EEG signals. These studies imply that substantial \emph{specific} information theoretically exists and is decodable from the signal which could potentially be utilized for communication with a non-invasive BCI. 

Deep learning techniques are eminently suited to the problem of extracting information from large and heterogeneous datasets. Specifically, foundation models \citep{bommasani2021opportunities} trained using predictive or reconstructive objectives over extremely wide datasets develop powerful and highly general representations \citep{gpt3,team2023gemini,rombach2022high}. These representations can then be utilized for a wide variety of downstream tasks. 

Our thinking is that if we can train fundamental foundation models to predict EEG signals, then these models will learn to extract a large fraction of the usable information within scalp-EEG signals which can then be applied to downstream tasks such as text or image prediction from EEG inputs, thus giving a direct route to noninvasive BCIs or, alternatively, ruling out this path if sufficient information simply does not exist within the EEG signal.

\subsection{Related Literature and Challenges}

A number of EEG foundation models have been proposed in recent years (e.g., \citealp{wang2023brainbert, cui2024neuro, zhang2024brant, yang2023biot, wan2023eegformer, jiang2024large, jiang2024neurolm, shi2024fome, wei2024transfer}), for a comprehensive review see \citet{kuruppu2025eeg}. Today, however, there are a number of challenges that face these EEG foundation models. Principally, the amount of publicly available EEG data is limited compared to other modalities such as image, video, or text. The scarcity of data renders brittle models, which struggle to generalize across datasets and recording conditions. Further, EEG datasets vary widely in the number of channels collected and the precise location of those channels. Many existing EEG models are trained on a fixed set of channel inputs and positions (a channel montage) and suffer limited generalizability across datasets. A second challenge unique to EEG is that the spatial layout of electrodes on the surface of the scalp, which influences correlations in the signals, is not reflected in the data structure, as it is for example in images or video. This renders common techniques such as CNNs or Transformers where attention is applied directly on sequence position ineffective.

\subsection{Contribution}

In this paper, we introduce \texttt{ZUNA} as a first step towards a general EEG foundation model. \texttt{ZUNA} is a 380M encoder-decoder model, able to take in sequences with arbitrary numbers and positions of channels, trained on both a regular and masked reconstruction objective. \texttt{ZUNA} can reconstruct existing channels using its encoder latent as a bottleneck, which forces it to learn general and powerful representations of EEG signals. If some channels are masked -- i.e. filled with zeros -- then \texttt{ZUNA} is also trained to reconstruct these channels from the information in the surrounding channels. This objective ensures that the model learns important cross-channel correlations and that the model is capable of directly predicting EEG signals from scratch. This masked channel reconstruction objective gives \texttt{ZUNA} useful capabilities such as denoising existing channels, reconstructing missing or noisy channels in realistic datasets, and upsampling the number of channels in a dataset, `imagining' new channels where none existed previously. \texttt{ZUNA} is trained on approximately two million channel-hours of publicly available scalp-EEG datasets.

Channel reconstruction and spatial upsampling are particularly valuable for researchers, clinicians, and BCI developers working with EEG data. A common problem in EEG data collection is that a channel may become noisy and potentially unusable for a time. This could be due to the electrode shifting position or temporary movements of the subject, which can suddenly inject noise into an otherwise clean channel. Concretely, given the remaining channels and a target electrode’s 3D scalp coordinate, the model generates a plausible time series for that location, enabling both infilling of missing or corrupted channels and spatial upsampling within a unified framework.

Additionally often cheaper or consumer grade EEG headsets do not come with the full complement of channels that commercial research headsets possess. This lower resolution makes it challenging to directly apply methods from the literature to these headsets without an unavoidable loss of accuracy. \texttt{ZUNA} provides the capability to ‘upsample’ and predict the outputs of channels at arbitrary locations on the head given the existing channels. While, clearly, the model is not infallible and relies only on the information it has access to, nevertheless, we show that our model substantially outperforms naive upsampling methods such as spherical-spline interpolation when the upsampling factor becomes large.

Our model makes the following contributions:
\begin{itemize}
    \item We introduce \texttt{ZUNA}, a novel diffusion-based encoder-decoder architecture with a novel and unique position embedding scheme designed to correctly represent the spatial information about channel location within EEG signals. This enables our model to generalize to novel channel positions instead of requiring a fixed channel schema, which enables applicability across many datasets and recording setups.
    \item We construct and harmonize a large-scale, heterogeneous EEG training corpus spanning 208 datasets with reliable 3D electrode metadata, comprising over 24 million 5-second samples and approximately 2 million channel-hours with up to 256 channels per sample.
    \item We demonstrate that the model consistently outperforms standard spherical-spline interpolation for channel reconstruction across multiple dropout regimes and evaluation datasets, with particularly strong gains at high upsampling factors.
    \item At only 380M parameters, our model is incredibly lightweight and can run extremely fast on any consumer GPU as well as decently on CPU. We provide the model weights under the highly permissive Apache 2.0 license and the weights are freely available on Huggingface \url{https://huggingface.co/Zyphra/ZUNA}, while inference and dataset preprocessing code is available on our GitHub \url{https://github.com/Zyphra/zuna} and as a pip package which is built on MNE python \url{https://pypi.org/project/zuna/}. We invite researchers and other EEG practitioners to experiment with the model for your own purposes and to send feedback to the corresponding authors which we can use to improve future iterations of the model. 
\end{itemize}

    
    \section{Data}

\begin{figure}[t]
    \centering

    \begin{subfigure}[t]{0.48\linewidth}
        \centering
        \includegraphics[width=\linewidth]{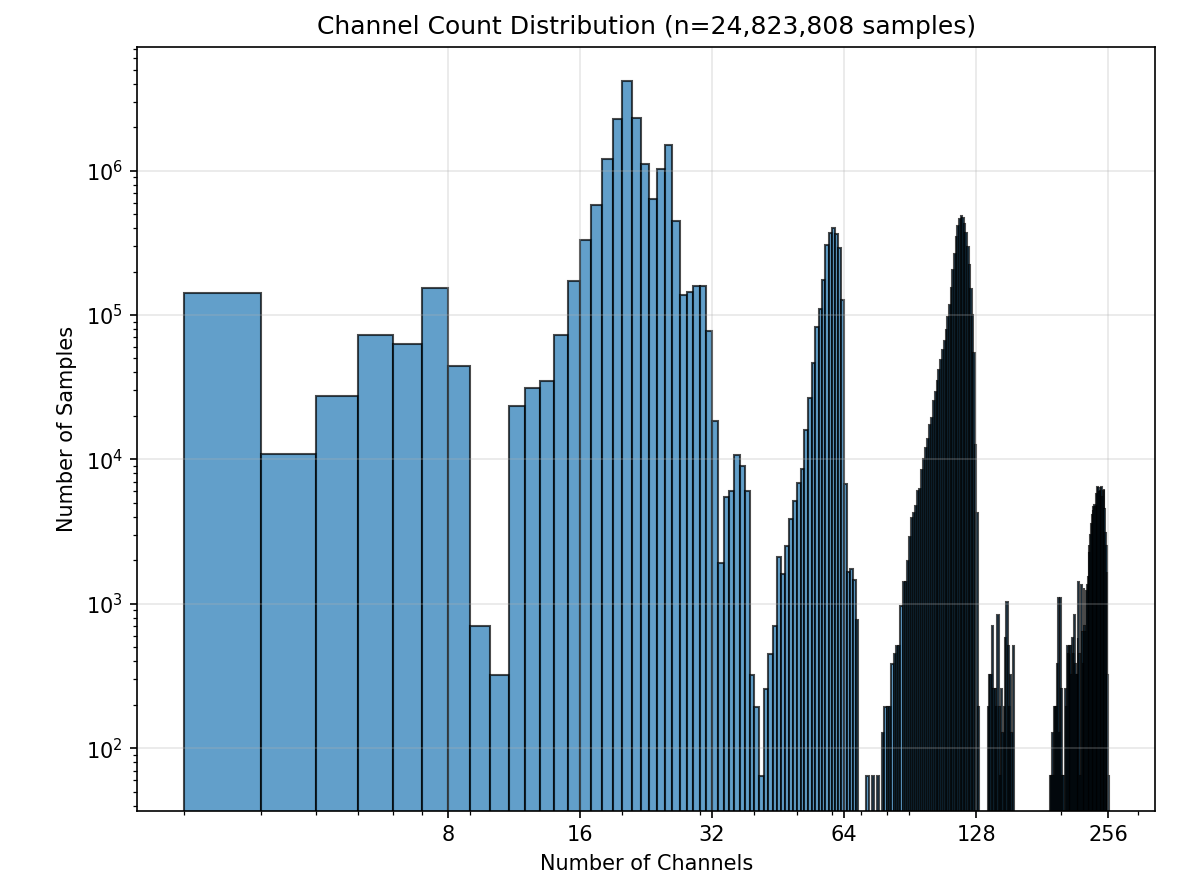}
        \caption{Channel count distributions after spatial filtering.}
        \label{fig:channel_distribution}
    \end{subfigure}
    \hfill
    \begin{subfigure}[t]{0.40\linewidth}
        \centering
        \includegraphics[width=\linewidth]{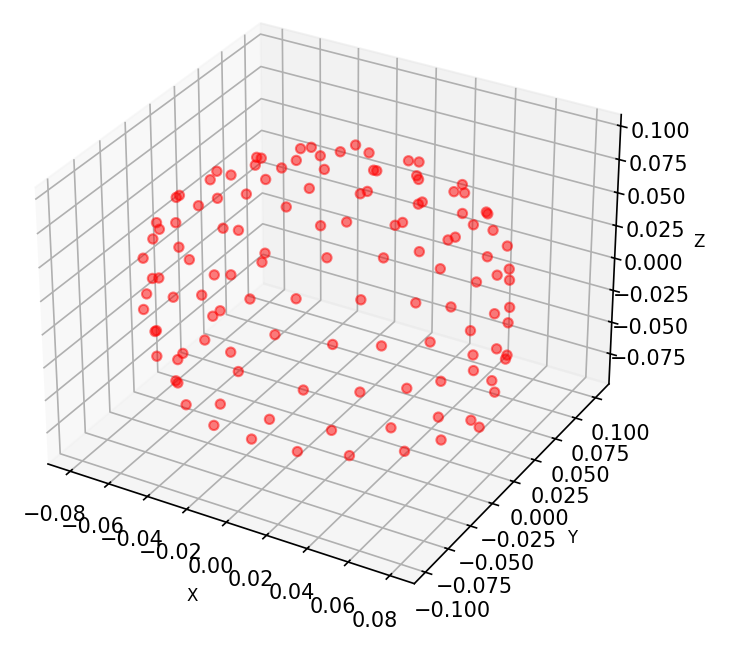}
        \caption{3D scalp electrode coordinates used for spatial conditioning.}
        \label{fig:eeg_3d_positions}
    \end{subfigure}

    \caption{
    \textbf{EEG channel configurations and spatial coverage in the training corpus.}
    \textbf{(a)} Distribution of retained EEG channel counts across all non-overlapping 5-second epochs after preprocessing, comprising approximately 2 million channel-hours with up to 256 channels per sample. The majority of samples have between 16–32 channels, with additional peaks at higher-density configurations (64, 128, and 256 channels). The x- and y-axis is shown on a logarithmic scale.
    \textbf{(b)} Three-dimensional Cartesian $(x,y,z)$ scalp coordinates of EEG electrode positions for a representative subset of 64 electrodes from the standard 10–20 montage, illustrating the spatial geometry used for conditioning the model.
    }
    \label{fig:channel_and_geometry}
\end{figure}
    
    \subsection{Data Preprocessing Pipeline}

To train a generalizable EEG foundation model across highly heterogeneous datasets, we constructed a unified preprocessing pipeline that standardizes sampling rate, temporal segmentation, normalization, and spatial metadata while deliberately preserving channel diversity across recordings. All preprocessing was implemented using MNE-Python and applied uniformly across datasets.

We aggregated data from two major open EEG repositories:
(i) the Temple University Hospital (TUH) EEG Corpus \citep{obeid2016temple}, and
(ii) a large collection of publicly available datasets hosted on OpenNeuro \citep{markiewicz2021openneuro}.
Together, these sources span a wide range of experimental paradigms, recording hardware, channel layouts, and data qualities.

Because \texttt{ZUNA} explicitly conditions on electrode location, only datasets and channels for which valid 3D scalp coordinates could be reconstructed were retained. Channel positions were inferred using a multi-stage strategy combining MNE standard montages (e.g., standard\_1005, EGI HydroCel layouts), BioSemi cap metadata, and manual dataset-specific mappings. After this spatial filtering step, the final corpus comprised 208 unique datasets for which we could derive reliable 3D scalp coordinates. Channels without resolvable 3D coordinates were excluded, resulting in the removal of an average of 5.82 channels per recording (median = 6). These primarily corresponded to ocular channels (e.g., EOG) and other auxiliary or non-standard sensors lacking well-defined scalp locations. No fixed channel subset or canonical montage was imposed, yielding a highly heterogeneous distribution of channel configurations across recordings as shown in \cref{fig:channel_and_geometry}.

All recordings were resampled to a common sampling rate of 256~Hz to ensure consistency across datasets. This included upsampling recordings originally acquired at lower sampling rates. Continuous recordings were segmented into non-overlapping 5-second segments (1280 samples per epoch). Recordings shorter than 10 seconds were discarded. To remove slow drifts, signals were high-pass filtered at 0.5~Hz and re-referenced to the common average reference. Line noise was detected adaptively on a per-recording basis by analyzing the power spectral density (Welch method) between 45~Hz and the Nyquist frequency. Narrowband spectral peaks exceeding a data-driven threshold were identified and notch-filtered, typically capturing 50~Hz, 60~Hz, and their harmonics where present. This adaptive approach avoids hard-coding line-noise frequencies and accommodates international datasets with differing power standards, which was necessary given our highly heterogeneous starting set of datasets.

To mitigate extreme artifacts while preserving maximal coverage, we applied a conservative signal-quality procedure at the channel and epoch level, where an epoch here denotes a non-overlapping 5-second segment of continuous EEG. Channels exhibiting near-flat signals (abnormally low variance relative to the recording-wide distribution, using a robust median/MAD criterion) or evidence of clipping (0.5\% of samples within a small tolerance of the channel minimum/maximum) were flagged as bad and subsequently zeroed out during epoch extraction. After high-pass filtering, re-referencing, and adaptive notch filtering, additional noisy channels (per 5s segment) as well as epochs were identified by excessive standardized variability (SD 3 per channel or epoch respectively) and removed. When more than 50\% of channels per epoch were marked as bad, the full epoch was discarded. 

After initial filtering, samples were then standardized using z-score normalization based on the global mean and standard deviation computed across all EEG channels within each recording, yielding approximately zero-mean, unit-variance signals while preserving relative spatial structure. For long recordings, normalization was performed independently on contiguous 10-minute segments; all subsequent analyses operated on non-overlapping 5-second epochs.

After spatial filtering and preprocessing, the final training corpus comprised 24,823,808 non-overlapping 5-second EEG epochs from a total of 208 datasets and approximately 2 million channel-hours of EEG recordings. The number of retained EEG channels per recording ranged from 2 to 256 (mean = 45.42; median = 22; SD = 41.50). The channel-count distribution was strongly multimodal (Figure\ref{fig:channel_and_geometry}, left), with the majority of recordings containing 16-32 channels and additional pronounced peaks at 64, 128, and 256 channels, reflecting common EEG acquisition systems.

After z-score normalization, signals within each recording were standardized. Each training sample corresponded to one 5-second segment sampled at 256Hz, yielding 1280 individual time points per channel. 

\section{Model}
     \begin{figure}
        \centering
        \includegraphics[width=0.9\linewidth]{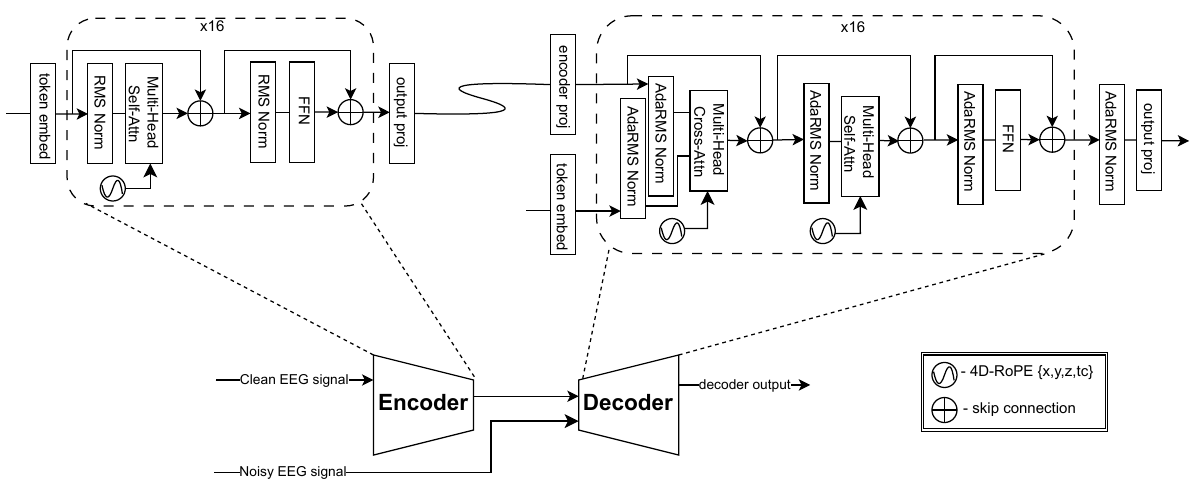}
        \caption{Diffusion Autoencoder EEG Model Architecture}
        \label{fig:placeholder}
    \end{figure}

\subsection{Diffusion models}

Diffusion models \citep{ho2020denoising,sohl2015deep} are widely used \emph{generative} models for continuous real-valued signals such as images and audio. They work by defining a noising process that gradually steps from clean data to pure noise \citep{karras2022elucidating,song2020score}. The model is then trained to invert this process, taking a step from noise towards data. Intuitively, diffusion models allow iterative denoising, giving the model many steps to converge to the correct signal, unlike a single step autoregressive model. Where well-behaved noising processes can be defined and inverted, diffusion models perform well, reaching state of the art performance in image and video generation \citep{nichol2021improved,dhariwal2021diffusion,rombach2022high,ho2022video} as well as performing strongly in audio modeling \citep{li2023starcoder,huang2022fastdiff}. Due to the continuous real-valued nature of EEG signals and the potential of their iterative denoising, for \texttt{ZUNA} we settled on a diffusion approach. 

Mathematically diffusion models define a Gaussian forward process that maps data to noise. Let $x_0 \sim p_{\text{data}}(x)$ denote a data point (e.g., an channel of EEG signals $\mathbb{R}^d$). We define the noising process as a Markov chain which adds a small amount of Gaussian noise at each step. 
\begin{align}
q(x_{1:T}\mid x_0)&=\prod_{t=1}^T q(x_t\mid x_{t-1}) \\ 
q(x_t \mid x_{t-1}) &= \mathcal{N}\!\bigl(x_t;\,\sqrt{1-\beta_t}\,x_{t-1},\,\beta_t I\bigr),
\end{align}
where $\beta_t$ is the noise level of each step. This noise schedule increases from close to $0$ to the standard deviation of the data during the noising process. Intuitively, early steps perturb fine details while preserving global structure, and late steps wash out semantics until $x_T$ is nearly isotropic Gaussian noise, making the terminal distribution easy to sample.

Conversely, generation is performed by learning an analogous reverse-time Markov chain
\begin{align}
p_\theta(x_{0:T})=p(x_T)\prod_{t=1}^T p_\theta(x_{t-1}\mid x_t)
\end{align}
with a simple prior $p(x_T)=\mathcal{N}(0,I)$. Crucially, due to the Gaussian dynamics, the \emph{posterior} $q(x_{t-1}\mid x_t,x_0)$ can be computed analytically. This lets us setup our learning objective as a principled and tractable variational lower bound,
\begin{align}
\log p_\theta(x_0)\ \ge\ \mathbb{E}_{q(x_{1:T}\mid x_0)}
\Big[\log p_\theta(x_{0:T})-\log q(x_{1:T}\mid x_0)\Big],
\end{align}
whose terms decompose across timesteps into KL divergences between Gaussians at each timestep. With an identity covariance the KL divergence simplifies to the standard mean-square-error objective. Classical diffusion models predict the injected noise $\varepsilon$ (or equivalently the score) with a neural network $\varepsilon_\theta(x_t,t)$, yielding the simple objective,
\begin{align}
\mathcal{L}_{\text{diff}}(\theta)
&=
\mathbb{E}_{t\sim \mathrm{Unif}\{1,\dots,T\},\,x_0}
\Big[\|\varepsilon - \varepsilon_\theta(x_t,t)\|_2^2\Big] \label{diffusion_objective} 
\end{align}
The key intuition is that learning to denoise at \emph{all} noise levels teaches the model a vector field that points back toward high-density regions of $p_{\text{data}}$. In the continuous-time limit this connects to score-based modeling \citep{hyvarinen2005estimation,vincent2011connection}, where the network approximates $\nabla_{x_t}\log q(x_t)$ and sampling follows a learned reverse diffusion (stochastic) or probability flow (deterministic) trajectory. In practice diffusion models are trained by sampling a minibatch of data at different levels of the noise schedule, and minimizing Eqn \ref{diffusion_objective} by gradient descent. While here we present the standard diffusion formulation for intuition, our implementation uses rectified flow \cite{liu2022rectifiedflow} parameterization for improved stability and sampling efficiency.

Diffusion autoencoders pair a regular diffusion decoder which maps noise to data, with an additional encoder which stores some information about the full signal in a latent bottleneck. This latent bottleneck can then be input into the decoder, typically through cross-attention to aid the denoising process. 

\subsection{\texttt{ZUNA} architecture}

Some prior works have applied deep learning models to the task of interpolating EEG signals or performing EEG "super resolution". Most similar to ours, \cite{wang2025stad} use diffusion in an encoder-decoder model where the encoder is convolution-based and the decoder is a diffusion transformer. \cite{sabasadiya2020dae} use a U-Net architecture based on CNNs with a fixed set of channels whose identity is determined by their location in the data array. \cite{li2025estformer} use a Masked Autoencoder based on a Transformer architecture with a fixed-masking strategy. \cite{corley2018eegsrgan} use an adversarial approach where both generator and discriminator networks rely heavily on CNNs. 
    
Prior works have typically assumed a fixed channel layout, fixed masking strategy, and even a specific fixed upsampling ratio — e.g. from 24 to 48 channels with always the same channels upsampled each time. Instead, we aim to produce a highly flexible model which can interpolate any number of dropped-out channels and where these channels can be placed at any point on the head. To achieve this, we utilized a novel 4-dimensional positional embedding system where we represented the x,y,z and time positions of each channel using a separate positional embedding scheme \cite{heo2024rope4vit}. This allows our model to flexibly generalize across channels at different locations and even interpolate entirely novel channel locations which were not recorded for a given signal. 
    
    \texttt{ZUNA} is a transformer-based encoder--decoder diffusion autoencoder for masked channel reconstruction of EEG signals. The model operates on a tokenized representation of multichannel EEG 2D structure (channel $\times$ time) while remaining compatible with standard 1D transformer processing. To achieve this, \texttt{ZUNA} introduces two architectural innovations:
\begin{enumerate}
  \item \textbf{Rasterized channel--time token ordering.} Each EEG channel is segmented into fixed-length windows ($0.125$~s, 32 samples) and each window is represented as a continuous-valued token. Tokens are computed from the raw time series through a token-encoder MLP. Tokens are then serialized in a raster-scan order: all channels for coarse time step 1, then all channels for coarse time step 2, etc. A 5~s sample with C channels contains $40C$ tokens. 
  \item \textbf{4D rotary positional encoding (4D RoPE).} Each token is associated with a 4D coordinate consisting of the channel's 3D scalp location and the token's coarse-time index. We inject this structure using a multi-dimensional RoPE variant following \cite{heo2024rope4vit}.
\end{enumerate}

Let the continuous EEG be $X \in \mathbb{R}^{C\times T}$, where $C$ is the number of channels. We segment each channel into non-overlapping (or fixed-stride) windows of length $\tau = 0.125$~s, yielding $M$ coarse windows per channel. Each channel--window pair is converted into a feature vector
\begin{align}
\mathbf{s}_{c,m} = \text{TokenEncoder}(X_{c,m})
\end{align}
We then serialize the 2D grid $(c,m)$ into a single 1D sequence of length $L = C\cdot M$ using a raster ordering:
\begin{align}
\bigl[\mathbf{s}_{1,1}, \mathbf{s}_{2,1}, \ldots, \mathbf{s}_{C,1},\ \mathbf{s}_{1,2}, \ldots, \mathbf{s}_{C,2},\ \ldots,\ \mathbf{s}_{C,M}\bigr].
\end{align}
This ordering is chosen so that (i) adjacent tokens share the same coarse-time index but different channels, and (ii) tokens at a fixed channel recur periodically every $C$ positions. 

Each token $i$ corresponds to a channel $c(i)$ and coarse time index $m(i)$. For positional embeddings, we associate a 4D discrete coordinate
\begin{align}
\mathbf{p}_i = \bigl(x_{c(i)}, y_{c(i)}, z_{c(i)}, m(i)\bigr),
\end{align}
where $(x_c,y_c,z_c)$ is the channel position on the head (discretized into forty `bins') and $m$ is the coarse-time (segment) index, $x_c,y_c,z_c,m \in \{1,50\}$. Note that 50 discretization bins was chosen because it was sufficient that each unique electrode in the training dataset had a unique bin once discretized. We then apply multi-dimensional rotary positional embeddings (RoPE) by passing the 4D index $\mathbf{p}_i$ into the rotary embedding module.

For the encoder and decoder models we use 16 transformer layers consisting of the standard self-attention plus MLP. Attention is bidirectional and we diffuse the entire set of channels at once during generation. To inject the encoder's latent information into the decoder we use adaptive-RMS norm, conditioned on the diffusion time step \cite{crowson2024adarmsnorm}. 

Within the encoder latent, following \cite{darcet2024vitregisters, jebraeeli2025vitcaeregisters}, we interleaved learnable register tokens which were inserted at a fixed stride. For a downsampling factor $d$, the token sequence is partitioned into groups of length $d$, and a learned register vector $r$ is prepended to each group:
\begin{align}
\bigl[\mathbf{r}, \mathbf{h}_1,\ldots,\mathbf{h}_d,\ \mathbf{r}, \mathbf{h}_{d+1},\ldots,\mathbf{h}_{2d},\ \ldots \bigr].
\end{align}
This produces an interleaved sequence of length $L' \approx L + L/d$. Following \citet{jebraeeli2025vitcaeregisters} and our prior experience in other modalities, we expect that the learned registers improve performance in practice although we did not specifically ablate this choice for EEG. In this work, we used a downsampling factor $d=1$. We set the sliding-window mask used in Flex Attention to 65536 tokens, a value large enough to include all tokens from a 5 second sample with 256 channels even after interleaving registers with $d=1$.

\subsection{Model Training}

To enable the model to infill signals on noisy, corrupted channels or to upsample to unobserved channels, we trained with a heavy channel dropout scheme. We set the probability of drop out to be 90\%. If dropout was selected, we randomly selected between 1 and C/2 channels to be dropped with 80\% probability or selected between C/2 and C-1 channels to be dropped with 20\% probability. During training, dropped out channels were replaced with zeros in the encoder input and targets (the noising direction) were sent to decoder input. During inference, the encoder receives zero inputs for dropped channels and the decoder only receives the random Gaussian initialization for both dropped and non-dropped channels. When we drop a channel we do not drop the channel's $\{x,y,z,t_c\}$ position and present this information to both encoder and decoder via 4D-RoPE. 
    
We trained with rectified flow loss \cite{liu2022rectifiedflow} and adaptive loss-weighting \cite{geng2024alw} on the decoder outputs and regularized the encoder latent with an auxiliary MMD loss \cite{dziugaite2015mmd}. The model was trained for 150000 steps with a learning rate of 1e-4 annealing down to 1e-6 with a cosine schedule. We used the AdamW optimizer with ($\beta_1$,$\beta_2$) = (0.9, 0.95). For training stability, we found that the dataset scale was incredibly important, and required us to adjust the dataset to have a standard deviation of $0.1$ with mean 0, and to adjust the noise variance of the diffusion process appropriately. Similarly, we observed that the global batch size also greatly impacted stability of training and required us to train with larger batches than expected. We trained with 6 gradient accumulation steps, yielding 90000 tokens per microbatch and 2.16M tokens per global batch across 3 nodes with 24 GPUs. Because of the highly variable numbers of channels per sample -- see Figure~\ref{fig:channel_distribution} -- naively padding the samples up to the maximum length would be extremely inefficient. To train efficiently, we implemented packing with sample-masking using flex attention to dynamically compute a special attention mask for each packed sample. Together, these improvements allowed us to train stably with high utilization at a rate of 50 samples per microbatch and 1150 samples per global batch. Training of the final model took approximately a week on three H100 nodes. Since the model size was so small, only data parallelism was used with large microbatch sizes. 

The spatiotemporal structure of EEG signals requires some modifications to the standard transformer architecture. Specifically, the transformer is only specifically designed to natively handle 1D sequences of elements while the EEG signal has $C$ channels in parallel. Our novel rotary embedding module means that the channels do not have to be specified in any specific order and enables the model to represent a variable numbers of channels. By training on heterogeneous datasets, the model instead learns to rely solely on the position embeddings which can handle arbitrary channels.

To validate our design choices, we performed a series of ablations on the model architecture and training scheme. First, we replaced the zeros-vector used to represent dropped out channels with a learnable vector that had norm similar to the data and input that into the encoder. After making the necessary changes to the inference code, we found that it performs comparably and save exploration in that direction for future work. We also performed an ablation on the best way to present the positional information to the model. We tried concatenating the $x,y,z$ position to the encoder input along with the fine-time EEG signal as part of the token while representing temporal position with standard 1D-RoPE, but this performed poorly. Finally, we performed an ablation testing whether sending the noised EEG signal into the decoder input for dropped channel improved performance relative to sending noise without the EEG signal. The relative performance in these two cases were similar.

\subsection{Usage}

Alongside this paper, we also release the model freely on huggingface with a permissive license which can be found at (\url{https://huggingface.co/Zyphra/ZUNA}). We also release inference code which can be found here (\url{https://github.com/Zyphra/zuna/}), and a pip package.

\section{Results}

\begin{figure}
    \centering
    \includegraphics[width=0.85\linewidth]{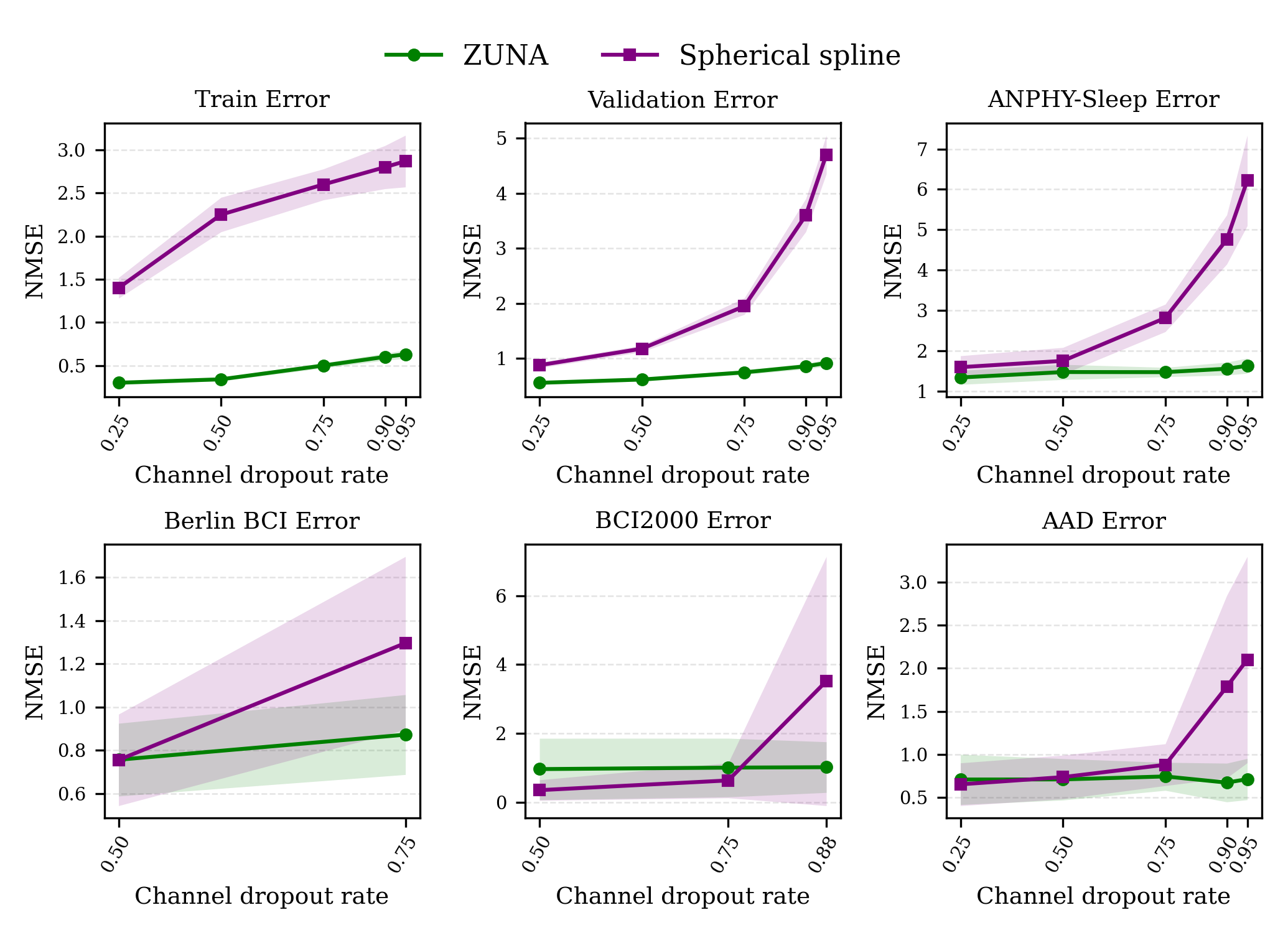}
    \caption{
    Normalized mean squared error (NMSE) for channel reconstruction as a function of channel dropout rate. Performance is shown for \texttt{ZUNA} (green) and spherical-spline interpolation (purple) under increasing levels of channel removal, corresponding to progressively more aggressive upsampling regimes. Results are reported for the training corpus (mixed channel configurations), a held-out validation set with a fixed 32-channel montage, the ANPHY-Sleep dataset (83 channels), Berlin BCI Competition III Dataset V (32 channels), the BCI2000 motor-imagery dataset (64 channels), and the ultra–high-density AAD dataset (255 channels). Across all datasets, reconstruction error increases with dropout rate for both methods, but \texttt{ZUNA} consistently outperforms spherical-spline interpolation, with the performance gap widening at higher dropout levels. Notably, the difference between methods is smallest for high-density recordings (e.g., 255 channels), where spatial interpolation is less challenging, whereas for lower-density montages and higher dropout rates, \texttt{ZUNA} maintains relatively stable performance while spherical-spline interpolation degrades substantially.}
    \label{fig:results}
\end{figure}

Since our model is primarily concerned with upsampling and interpolating channels, we take reconstruction fidelity as our primary evaluation metric. To assess both in-distribution performance and cross-dataset generalization, we validated \texttt{ZUNA} on the training corpus, a held-out validation set with a fixed 32-channel montage, and four additional publicly available EEG datasets not included in the training sources (OpenNeuro and TUH). These held-out datasets were selected to probe generalization across recording paradigms, channel densities, and spatial layouts. 

First, we used a subset of the ANPHY-Sleep dataset \citep{wei2024anphy}, an open high-density polysomnography resource comprising overnight EEG recordings from healthy adults with 83-channels being recorded. In our experiments we used data from the first 10 participants and restricted the analysis to the first 30 minutes of recording during which participants were still awake. 

Second, DatasetV of the Berlin BCI CompetitionIII \citep{millan2004need} is a widely used motor-imagery EEG benchmark recorded with a moderate-density scalp montage. This dataset represents a classic low-channel-count BCI setting and enables comparison with prior EEG channel-reconstruction and superresolution models that report results on this benchmark.

Third, we evaluated the model on an EEG dataset recorded using the BCI2000 platform \citep{schalk2004bci2000}, a widely used general-purpose system for brain–computer interface research. The dataset used here consists of 109 participants performing a series of motor execution and motor imagery tasks recorded with a 64-channel EEG montage. This dataset represents a typical medium-density motor-imagery BCI scenario with a large and diverse subject pool.

Finally, we used the Ultra-high density 255-channel EEG auditory attention decoding (AAD) dataset \citep{mundanad2021eeg}, a publicly released collection of multi-subject EEG data recorded with an international 10-5 electrode array at 255 channels while subjects attended to competing auditory streams. From this dataset we included all available EEG recordings from the first five participants in our evaluation.

We take spherical-spline interpolation \citep{perrin1989sphericalspline} between channels as our baseline, which is widely used by practitioners and is implemented in common EEG processing packages such as MNE \citep{gramfort2013meg}. Spherical-spline interpolation is a surprisingly strong baseline for scalp-EEG signals, since volume conduction and skull conductivity induce a strong spatial smoothness, such that nearby electrodes share substantial linear signal components. 

We evaluated reconstruction performance under varying degrees of channel dropout. These rates correspond one-to-one with the commonly reported ``scale factor'' of the superresolution literature. For example, in the ANPHY-Sleep dataset we randomly removed 20\%, 50\%, 75\%, and 90\% of channels at inference time, corresponding to upsampling factors of 1.25$\times$, 2$\times$, 4$\times$, and 10$\times$, respectively. For each condition, the model was tasked with reconstructing the dropped channels from the remaining observed channels, and performance was compared against spherical-spline interpolation.

Across all evaluation datasets, our model outperformed MNE's inbuilt and ubiquitous spherical-spline interpolation method and clearly pulled ahead at higher dropout rates. The performance gap increased systematically with higher dropout rates, with the largest gains observed in the most extreme (90\%) dropout condition. This divergence was driven primarily by a sharp degradation in MNE's interpolation performance as channel density decreased, whereas our model retained substantially higher reconstruction fidelity even when only a small subset of channels was available. 

Notably, this pattern held across both low- to mid-density and ultra-high density recording regimes, indicating that the model generalizes across channel layouts and does not rely on assumptions specific to a particular montage. Together, these results suggest that learning spatially grounded representations over heterogeneous EEG data enables robust channel reconstruction well beyond what is achievable with purely geometric interpolation methods.

It should be noted that several evaluation-specific factors likely contributed to dataset-dependent performance differences. A practical limitation of our evaluation is that some benchmark datasets (notably BCI2000 and Berlin BCI Competition III Dataset V) are distributed primarily as pre-segmented, stimulus-locked epochs rather than continuous raw recordings. Because our training pipeline is built around continuous data (to enable robust line-noise detection and stable filtering), we adapted these epoch-based datasets by concatenating epochs to estimate line-noise peaks and apply high-pass filtering in a manner consistent with training, before re-segmenting and applying notch filtering at the epoch level. This workflow is necessarily suboptimal: filtering short epochs can introduce edge transients and boundary artifacts (especially for IIR/FIR operations near the segment boundaries), and concatenation/re-segmentation can yield filtering behavior that differs from true continuous preprocessing. We suspect this preprocessing mismatch contributes to the comparatively weaker performance observed on these datasets. In practice, we recommend applying \texttt{ZUNA} to continuous raw EEG whenever possible (or at least to long continuous segments) to avoid short-epoch filtering artifacts and to more closely match the model’s training distribution.

While there are other similar models in the literature, we found that very few of them released either the model weights or complete inference code, making direct fair quantitative comparisons difficult and reproduction of their results effectively impossible. Moreover, reported reconstruction metrics (e.g., MSE or NMSE) are highly sensitive to preprocessing choices, channel layouts, and evaluation protocols, and therefore specific values of these metrics cannot be compared across studies. To address this limitation, we release the trained model weights alongside a fully documented inference pipeline and a lightweight \texttt{pip-}installable package that integrates directly with common EEG processing workflows (e.g., MNE-Python). This allows practitioners to reproduce our results, apply the model to their own data, and perform controlled comparisons under matched preprocessing and evaluation settings.

\section{Discussion}

In this paper, we have introduced \texttt{ZUNA}, a small but highly capable channel infilling model. We have demonstrated our model substantially outperforms ubiquitous methods of infilling that practitioners and researchers use today. Moreover, our model performs increasingly better as the difficulty of the task increases as we drop out more and more channels. Crucially, unlike other models in the literature, our model generalizes across datasets and across channel number and positions, potentially allowing it to be flexibly applied to a wide range of EEG datasets in the wild. One note of caution, however, is that like all generative models, our model is trained to generate plausible continuations given its limited observations, which means that it can \emph{hallucinate} incorrect EEG signals. Channels reconstructed by our model should be treated as imputed data and not ground-truth measurement particularly in clinical decision-making.

The performance of spherical-spline interpolation depends strongly on montage density. For ultra–high-density recordings (e.g., the 255-channel AAD dataset), neighboring electrodes are closely spaced and scalp EEG is spatially smooth due to volume conduction and skull-related blurring; under these conditions, geometric interpolation is a strong baseline, particularly at low-to-moderate dropout where the missing channels remain well constrained by nearby measurements. This explains why spline interpolation performs very well in the low-dropout regime on the 255-channel AAD  dataset and why the absolute gap between methods is smaller there. In contrast, as channel density decreases and/or dropout becomes more aggressive, the spatial gaps between observed electrodes grow and spline interpolation degrades sharply because it has no learned prior beyond smoothness. \texttt{ZUNA} is most advantageous in precisely these low-density / high-upsampling settings, where it can leverage learned multivariate structure across channels and datasets to generate plausible reconstructions that go beyond purely geometric smoothing.

While our model achieves an impressive level of performance and generality, demonstrating the power of deep learning methods as seen across other domains, there are still many potential avenues for improvement.

Firstly, our model and data scales are both very small compared to other, more mature, modalities such as audio, vision, and text. We see the preliminary signs of classical scaling-law-like trends in EEG as well and this would point to substantial further improvements being possible simply by scaling the model and data sizes. In terms of model size, there is clearly ample room to scale. Data is the more challenging axis to scale. To train \texttt{ZUNA}, we utilized existing open-source datasets, however unlike other modalities, there is nowhere near as much EEG data available on the open web. Scaling up the available and open datasets is thus an important and general challenge for the field. 

Beyond simply scale, there are also various avenues of improvement available. One method, which also somewhat helps ameliorate the data scaling challenges, is to also include intracortical EEG. There is some evidence that there is transfer between intracortical and scalp-EEG \citep{yuan2024brainwave}, and the general expectation from applying deep learning in other modalities, is that integrating new, but similar, areas of data induces transfer and improves modeling capabilities generally. 

A second clear avenue for improvement is the context length of the model. For now, our model has been trained to process sequences only up to five seconds in length, however full EEG recordings can last many minutes to hours. We aim to apply techniques on context length extension and general long-context training from other well-developed areas of deep learning, especially learning from text data, to create models which can meaningfully learn from and interact with substantially longer EEG time series. 

Additional avenues are improvements to the core architecture of the model. Our model attempts to do triple duty as a compressor to continuous tokens for consumption by downstream models, a model capable of reconstructing dropped out channel signals, and a model with powerful and general latent representations of EEG in its encoder bottleneck, which can then be built upon by classifiers. Certainly designing specialized architectures and learning objectives for these tasks may help improve performance at a given level of model and data scale. Secondarily, our architecture is primarily adapted from those used for audio autoencoders and has not particularly been designed with the specific characteristics of EEG signals in mind, except for our innovations about representing the positional information and channels. Moreover, our model only takes in the time-domain signal. Adjusting the model architecture to better fit the natural characteristics of the EEG signal and performing input feature engineering could further improve performance.

Beyond the utility of EEG foundation models for practitioners, we believe they also hold deep scientific interest. There are fundamental open questions regarding the fraction of useful neural information that can be captured by scalp-EEG. Large-scale foundation models excel at extracting and storing the information from a massive and messy corpora of data in their weights.

By training foundation models, we obtain artifacts that can be explored and mechanistically understood and which can provide evidence about the presence (or absence) of specific information within the EEG signal, about brain states and about correlates with behavior. Moreover, determining the extent to which many behaviors or mental states can be inferred from purely scalp-EEG signal also provides significant information about the feasibility and utility of non-invasive BCI in general. 

One final note is that throughout we have striven to make this model accessible and useful to EEG researchers and other practitioners in the field. This includes making the model weights fully open source and easily accessible on hugging face, and providing straightforward and usable inference code and EEG preprocessing code for the model.  We plan to continue improving the accessibility and utility of our models in future releases with an eye towards improving integrations with commonly used software tooling.

\clearpage

\bibliographystyle{tmlr}
\bibliography{main}

\clearpage


\newpage
\section{Appendix}



    \begin{figure}[H]
        \centering
        \includegraphics[width=0.95\linewidth]{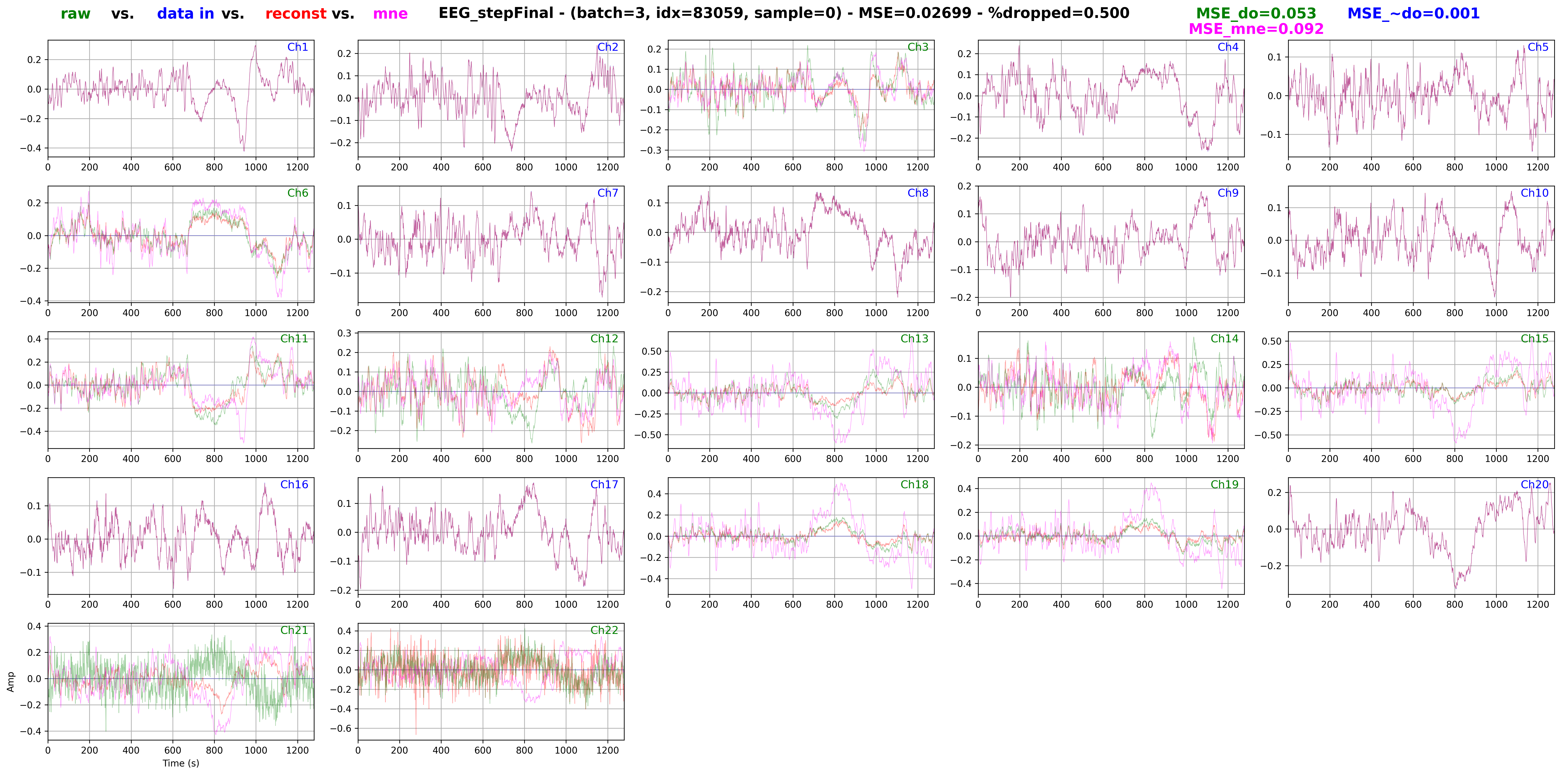}
        \caption{Full sample reconstruction of a 22-channel sample from Figure \ref{fig:Reconstructions}}
        \label{fig:Reconstructions full}
    \end{figure}

\end{document}